 
\baselineskip15.5pt
\abovedisplayskip 4pt plus3pt minus2pt
\belowdisplayskip 3pt plus3pt minus2pt
\vsize=9.7 true in \raggedbottom
\hsize=6.5 in
\voffset= -0.2 true in
\font\bmit=cmmib10 \textfont9=\bmit \def\bmit{\fam9 }
\font\bx=cmbsy10 \textfont10=\bx \def\bx{\fam10 }

\mathchardef\beta="710C
\mathchardef\pi="7119
\mathchardef\sigma="711B
\mathchardef\lambda="7115
\mathchardef\mu="7116
\mathchardef\nabla="7272
\parskip=6pt plus3pt
\def\boxit#1{\vbox{\hrule\hbox{\vrule\kern1pc
   \vbox{\kern1pc#1\kern1pc}\kern1pc\vrule}\hrule}}
\def\sqr#1#2{{\vcenter{\vbox{\hrule height.#2pt
   \hbox{\vrule width.#2pt height#1pt  \kern#1pt
      \vrule width.#2pt}
     \hrule height.#2pt }}}}

 \leftline{\bf NON-DISPERSIVE WAVEPACKET SOLUTIONS OF THE
 SCHRODINGER EQUATION
 \hfil\break  }

\rightline{Shaun N Mosley,\footnote{${}^*$} {E-mail:
shaun.mosley@ntlworld.com  }
Sunnybank, Albert Road, Nottingham NG3 4JD, UK }

\beginsection Abstract

 The free Schrodinger equation has constant velocity wavepacket
solutions $ \psi_{\bf v} $ of the form
$ \psi = f ({\bf r} - \, {\bf v}t ) \,
e^{ - \, i \, m c^2 t / 2 }.$ These solutions
are eigenvectors of a momentum operator $ {\bf \tilde p} \, $
which is symmetric in a positive definite scalar product space.
 We discuss whether these $ \psi_{\bf v} $ can act as
 basis states rather than the usual plane wave solutions.
 

\beginsection 1 Introduction

 We establish
 a class of solutions to the Schrodinger equation that are
 localised, their amplitude inversely proportional
 to the distance from the wavepacket centre; they are
 non-dispersive, and are eigenfunctions of a particular
 momentum operator found below. We show that the Schrodinger
 equation
 $$  ( i \, \partial_t \, + \, { \nabla^2 / \, 2 \, m }  ) \,
 \psi  = 0 \,   \eqno (1)  $$
(we put $\hbar = 1 \, $ throughout )
admits the following bounded solution representing a wavepacket
travelling with velocity $ v \, $ in the $ z \, $ direction
$$ \eqalignno{
\psi_{ v }
 &= \, { \sin \Big( m (c^2 + v^2)^{1/2} \; ( x^2 + \, y^2 + \,
  (z - v \, t )^2 )^{1/2} \Big) \over ( x^2 + \, y^2  + \,
  (z - v \, t )^2 )^{1/2} } \,
   e^{  i \, m \, v \, (z - v t ) }  \,
   e^{ - \, i \, m c^2 t / 2 } , \,  \cr
   }  $$
  or more generally with velocity $ {\bf v} \, :$
 $$ \eqalignno{
 \psi_{{\bf v}}
&= \, { \sin ( m v^0 | {\bf r} \, - \, {\bf v} t | )
 \over  | {\bf r} \, - \, {\bf v} t |  } \,
 e^{ i \, m {\bf v} \cdot ({\bf r} \, - \, {\bf v} t ) } \,
 e^{ - \, i \, m c^2 t / 2 }  & (2) \cr
 &\equiv \, m v^0 j_0 ( m v^0 {\bar r} )  \,
  e^{ i \, m  {\bf v} \cdot {\bf \bar r} } \,
  e^{ - \, i \, m c^2 t / 2 }        \cr
     } $$
with
$$  v^0 \equiv (c^2 + |{\bf v}|^2)^{1/2} = \, (c^2 + v^2)^{1/2} ,
 \qquad  {\bf \bar r} \equiv  {\bf r} \,
 - \, {\bf v} t  , \qquad \bar r \equiv | {\bf \bar r} | \, ,$$
 and $ j_0 \, $ is the spherical Bessel
 function of order zero. The $ c \, $ is an arbitrary
 parameter having dimension velocity: it can take any finite
 value or be zero, as in an earlier
 version of this paper. From now on we will consider
 $ c \, $ to the speed of light then
 $$ m c v^0 = \, m c \, (c^2 + v^2)^{1/2} \, \approx \,
 m c^2 + \, \textstyle{1 \over 2} m \, v^2  \qquad \qquad
 \hbox{for } v \ll c \, , $$
 which is the non-relativistic energy after subtracting out the
 `rest energy' $ m c^2 .$ In the next section we
 find an operator $ \tilde{p}^0 $ such that
  $ \tilde{p}^0 \psi_{\bf v} = \, m v^0 \, \psi_{\bf v} \, .$

 To verify that (2) is a solution of (1), note the identities
  $$ \eqalignno{
  &( \partial_t + {\bf v} \cdot {\bx \nabla} \,
   + { i \, m c^2 \over 2 } ) \,
\psi_{{\bf v}} \, = \, 0 & (3) \cr
&( {\bx \nabla } \, - \, i \, m \, {\bf v} ) \, \psi_{\bf v} \,
 = \, e^{  i \, m  {\bf v} \cdot ( {\bf r} \,
 - \, {\bf v} t ) } \, e^{ - \, i \, m c^2 t / 2 } \; \,
  {\bx \nabla } \, \Big[{ \sin ( m v^0 | {\bf r} \,
  - \, {\bf v} t | )
\over  | {\bf r} \, - \, {\bf v} t |  } \Big] \,  \cr
   &( {\bx \nabla } \, - \, i \, m \, {\bf v} )^2 \,
   \psi_{\bf v} \, = \, e^{  i \, m  {\bf v} \cdot ( {\bf r} \,
  - \, {\bf v} t ) } \, e^{ - \, i \, m c^2 t / 2 } \; \,
  \nabla^2 \, \Big[{ \sin ( m v^0 | {\bf r} \,
  - \, {\bf v} t | )
\over  | {\bf r} \, - \, {\bf v} t |  } \Big]  \,
 = \, - \, m^2 {v^0}^2 \, \psi_{\bf v} \, . & (4)  \cr
   }  $$
The first identity (3) follows from $ \psi_{{\bf v}}
 = f ( {\bf r} \, - \, {\bf v} t ) \,
 e^{ - \, i \, m c^2 t / 2 } \, .$ Expanding out (4)
 and substituting in (3) we obtain
   $$ \eqalignno{
  &\Big( \nabla^2 \, - \, 2 \, i \, m \,
  {\bf v} \cdot {\bx \nabla} \,
   - \, m^2 v^2 \Big)  \, \psi_{\bf v}  \,
   = \, - \, m^2 ( c^2 + v^2 ) \, \psi_{\bf v} \cr
  &\Big( \nabla^2 \, + \, 2 \, i \, m \, ( \partial_t
  + { i \, m c^2 \over 2 } ) \, \Big) \,
    \psi_{\bf v} \,  = \, - \, m^2 c^2 \, \psi_{\bf v} \cr
  &\Big( \nabla^2 \, + \, 2 \, i \, m \, \partial_t \Big) \,
    \psi_{\bf v} \,  = \, 0 \cr
   }  $$
   which is (1).

   That $ \psi_{{\bf v}}
  = \, f ( {\bf r} \, - \, {\bf v} t ) \,
  e^{ - \, i \, m c^2 t / 2 } \, $ is a clear statement of its
  non-dispersive property, the $ e^{ - \, i \, m c^2 t / 2 } \, $
  being merely a time phase factor over all space.
  The wavepacket centre is at $ {\bf r} = {\bf v} t \, $.
 The only free parameter in $ \psi_{\bf v} \, $ is
  $ {\bf v} \, $ itself (once we have fixed the value of
  $ c \, ),$ which makes us
curious as to whether  $ \Psi_{\bf v} \, $ is an eigenfunction.
We will go on to consider whether we can regard these $
\psi_{\bf v} \, $ as the basis states of the
Schrodinger equation, rather than the usual plane wave solutions
which are spread out over all space.

 I have not found any reference to the $ \psi_{\bf v} \, $
  solutions in the literature: it is sometimes stated that
 such constant velocity non-dispersive wavepackets do not exist
 (see for example p167 of ref [1]). Wavepacket
 solutions of the Schrodinger equation are considered in
 Besieris et al [2]
 (see (2.7) therein), and Barut [3]. A constant acceleration
 wavepacket [4], sometimes called the Airy packet, is well
 known.

 \beginsection 2 The velocity (momentum) operators

  The momentum of a particle is simply the product of its mass
  and velocity $ {\bf p} = \, m \,{\bf v} $
   as usual for the non-relativistic case. As
   $ v^0 = \sqrt{ c^2 + \, v^2 } \, $ we will label
 $$ p^0 = m v^0 = \sqrt{ m^2 c^2 +  p^2 } \, .$$
 The $ p^0 $ has relativistic character, even though the
 absolute velocity has no upper limit. In terms of
 $ {\bf p} \, $ the wavefunction (2) is
 $$ \eqalignno{
 \psi_{{\bf v}}
&= \, { \sin ( p^0 | {\bf r} \, - \, {\bf p} t / m | )
 \over  | {\bf r} \, - \, {\bf p} t / m|  } \,
 e^{ i \,  {\bf p} \cdot ({\bf r} \, - \, {\bf p} t / m ) } \,
 e^{ - \, i \, m c^2 t / 2 } \, , \cr
 \noalign{\noindent and at time  $ t = 0 \, $ }
 \psi_{{\bf v}} &\rightarrow \, {1 \over r }  \,
 \sin ( p^0 \, r ) \,
 e^{ i \, {\bf p} \cdot {\bf r} } \,
 = \, - {i \over 2 \, r } \, \Big[ e^{i \,  p^0 \, r \,
 + i \, {\bf p} \cdot {\bf r} } - \,
 e^{ - \, i \,  p^0 \, r \,
 + i \, {\bf p} \cdot {\bf r} } \Big] \,. & (5)  \cr
     } $$
 
     We first establish the
 following identities, which can be verified by direct
 calculation:
   $$ \eqalignno{
 \textstyle{1 \over 2} \,
  r \, ( - m^2 c^2 + \nabla^2 ) \, ( e^{ \pm i \,  p^0 \, r }
  e^{ i \, {\bf p} \cdot {\bf r} } )
 &=  \, \pm \, i \, ( \partial_r r ) \,
  p^0 \, ( e^{ \pm i \, p^0 \, r \, }
  e^{ i \, {\bf p} \cdot {\bf r} } )  & (6) \cr
  {\bx \nabla}  \, ( e^{ \pm i \,  p^0 \, r \, }
  e^{ i \, {\bf p} \cdot {\bf r} } )
 &=  \, ( \, i \, {\bf p} \, \pm \, i \,
  p^0 \, {\bf \hat r} ) \; ( e^{ \pm i \,  p^0 \, r }
  e^{ i \, {\bf p} \cdot {\bf r} } )  \cr
  ( \partial_r r ) \, {\bx \nabla}  \,
  ( e^{ \pm i \, p^0 \, r \, }
  e^{ i \, {\bf p} \cdot {\bf r} } )
 &=  \, ( \, i \, {\bf p} \, \pm \, i \,
  p^0 \, {\bf \hat r} ) \; ( \partial_r r ) \,
  ( e^{ \pm i \,  p^0 \, r \, }
  e^{ i \, {\bf p} \cdot {\bf r} } ) & (7) \cr
 \noalign{\noindent now we multiply (6) by $ {\bf \hat r}
 \equiv \, {\bf r} / r \, $
 and subtract (7) obtaining }
 [ \, - \, ( \partial_r r ) \, {\bx \nabla } \,
   - \, \textstyle{1 \over 2} \,
 {\bf r} \, ( m^2 c^2 - \, \nabla^2 ) ] \; ( e^{ \pm i \,
  p^0 \, r \, }
  e^{ i \, {\bf p} \cdot {\bf r} } )
  &=  \, - \, i \, ( \partial_r r ) \,
   {\bf p} \, ( e^{ \pm i \,  p^0 \, r \, }
  e^{ i \, {\bf p} \cdot {\bf r} } )      \cr
 [ \, - \, ( \partial_r r ) \, {\bx \nabla } \,
   - \, \textstyle{1 \over 2} \,
  {\bf r} \, ( m^2 c^2 - \, \nabla^2 ) ] \, \{\sin ( p^0 \, r )  \,
  e^{  i \,  {\bf p} \cdot {\bf r}  }  \}
 &=  \, - \, i \, ( \partial_r r ) \,
   {\bf p} \, \{ \sin ( p^0 \, r )  \,
  e^{  i \,  {\bf p} \cdot {\bf r} }  \}    \cr
  ( - \, \textstyle{1 \over 2} \, m^2 c^2 {\bf r} \,
  + \, {\bf a} ) \, r \, \psi_{\bf v} \,
 &=  \, - \, i \, ( \partial_r r ) \, r \, {\bf p} \,
  \psi_{\bf v} \, .     & (8)    \cr
     } $$
 The operators $ {\bf a} \, ,$ and $ a^0 $ with
 $ (a^0)^2 = {\bf a}^2  ,$ are
    $$ ( a^0 , \, {\bf a} ) \,
  = \, ( - \, \textstyle{1 \over 2} \,
 r \, \nabla^2 \, , \; - \, ( \partial_r r ) \,
 {\bx \nabla } \, + \, \textstyle{1 \over 2} \,
 {\bf r} \, \nabla^2 ) \, . \eqno (9) $$
  The operator $ {\bf a} \, $
 is related to the Runge-Lenz operator used for solving
  the Schrodinger equation with a Coulomb potential,
 its components commute with each other. Further properties
 of $ ( a^0 , \, {\bf a} ) \, $ are listed in [5].
 
 We now introduce the dilation operator $ \Sigma \, $
  $$ \eqalignno{
 &\Sigma \equiv \, - \, i \, ( \partial_r r ) \, \cr
  \noalign{\noindent and an inverse
  dilation operator $ \Sigma^{-1} \, :$ }
 &\Sigma^{-1} \, f (r , \theta , \phi ) \,
  = \, {i \over r }  \, \int_0^r f (t , \theta , \phi ) \, d t \,
  , \qquad \qquad \Sigma^{-1} \, f ({\bf r} ) \,
  = \, i \, \int_0^1 f (\lambda {\bf r} ) \, d \lambda \, .
   & (10) \cr
    }    $$
  The operators $ \Sigma, \, \Sigma^{-1}  \, $
  commute with any operator of homogeneity degree zero,
  i.e. any operator such as $ ( r {\bx \nabla } ) \, $
  which is invariant under a dilation of $ {\bf r} \, .$
 Now we can write (8) as
 $$ \eqalignno{
 & ( - \, \textstyle{1 \over 2} \, m^2 c^2 {\bf r} \,
  + \, {\bf a} ) \,  \, r \, \psi_{\bf v} \,
 =  \, \Sigma \, r \, {\bf p} \, \psi_{\bf v} \,  \cr
 \noalign{\noindent
   and multiplying from the left by  $ {1 \over r} \,
   \Sigma^{-1} \, $ we obtain }
 &{\bf \tilde p} \, \psi_{\bf v} \,
  \equiv  \, \textstyle{1 \over r} \, \Sigma^{-1} \,
   ( - \, \textstyle{1 \over 2} \, m^2 c^2 {\bf r} \,
  + \, {\bf a} ) \, r \, \psi_{\bf v} \,
  =  \, {\bf p} \, \psi_{\bf v} \, ,   \cr
 &{\bf \tilde p} \,
  \equiv  \, \textstyle{1 \over r} \, \Sigma^{-1} \,
   ( - \, \textstyle{1 \over 2} \, m^2 c^2 {\bf r} \,
  + \, {\bf a} ) \, r \,   & (11)  \cr
     } $$
 which is the momentum operator.

 \noindent{\it The operator } $ \tilde{p}^0 \, $
 \hfill\break \noindent
 We look for an operator $ \tilde{p}^0 $ such that
 $ \tilde{p}^0 \, \psi_{\bf v} = \, p^0 \; \psi_{\bf v} \,
  ,$ a more difficult task than finding $ {\bf \tilde p} \, .$
 As the notation implies, we will show in the next section that
 $ ( p^0 , \, {\bf \tilde p} ) \, $ is a 4-vector, finding
 the boost operator generating a change in momentum.
 From (6) we find that
      $$ \eqalignno{
      (  \textstyle{1 \over 2} \, m^2 c^2 r \,
  + \, a^0 )  \, \{\sin ( p^0 r )  \,
  e^{ i \, {\bf p} \cdot {\bf r} }  \} \,
 &\equiv \, \textstyle{1 \over 2} \,
  r \, ( m^2 c^2 - \, \nabla^2 ) \, \{\sin ( p^0 r )  \,
  e^{ i \, {\bf p} \cdot {\bf r} }  \}
 = \, - \, ( \partial_r r ) \, p^0 \{ \cos ( p^0 r )  \,
  e^{ i \, {\bf p} \cdot {\bf r}  }  \} \, \cr
 &                                  & (12) \cr
    }  $$
 which cannot immediately be resolved into an eigenvalue equation
due to the cosine function on the RHS instead of a sine function.
In Ref [6] we constructed a Hilbert transform operator
$ {\cal H}_- $ which
converts the $ \{ \cos ( p^0 \, r )  \,
  e^{ i \, {\bf p} \cdot {\bf r} }  \} \, $ into
 $ \{ \sin ( p^0 \, r ) \, e^{ i \, {\bf p} \cdot {\bf r} } \}
 \, .$ This $ {\cal H}_- $ is defined
$$ \eqalignno{
{\cal H}_\pm  f({\bf r})
&\equiv \, {1 \over 2} \, \left[ \, ( {\cal H}_e + {\cal H}_o )  \,
 \pm \,  ( {\cal H}_e - {\cal H}_o )  \, {\cal P} \, \right] \,
 f({\bf r}) &  (13) \cr
&= \,  { 1 \over \pi } \, \int_0^\infty
 \Big( \,  {  f( \lambda {\bf r}) \over \lambda - 1 } \,
 - \, {  f( - \lambda {\bf r}) \over 1 + \lambda  } \, \Big) \;
 d\lambda \, \qquad  \cr
     }  $$
     where $ {\cal H}_e , \, {\cal H}_o \, $ are the Hilbert
     transforms of even, odd functions:
$$ \eqalign{
{\cal H}_e \, f({\bf r}) = \, - \, { 2  \over \pi } \,
\int_0^\infty { \, f( \lambda {\bf r} ) \over 1 - \lambda^2 } \,
d\lambda \, , \qquad
{\cal H}_o \, f({\bf r}) =  \, - \,  { 2 \over \pi } \,
\int_0^\infty { \lambda \, f( \lambda {\bf r} ) \over 1 - \lambda^2 } \,
d\lambda \, . \cr }
  \eqno (14)   $$
  Note that $ {\cal H}_\pm $  involves an integration over the
   entire axis $ {\bf r} =  \lambda \, {\bf \hat{r}} \, ,\;
( - \infty < \lambda < \infty ) $ through the origin to the
field point.
We list some properties of these operators
(for details see [6]) to be used later, which may be verified
from the definitions (13) and (14):
$$ \eqalignno{
 &{\cal H}_e {\cal H}_o = {\cal H}_o {\cal H}_e = \, - \, 1 \,
  \quad \; \Rightarrow \qquad {\cal H}_- {\cal H}_+
 = {\cal H}_+ {\cal H}_- = \, - \, 1 \, , & (15) \cr
 &\partial_r {\cal H}_\pm = \,  {\cal H}_\mp \, \partial_r \, ,
 \qquad \qquad {\cal H}_\pm r \, = \, r \, {\cal H}_\mp \,
 \qquad \qquad {\cal H}_\pm {\bf \hat r} \,
 = \, {\bf \hat r} \, {\cal H}_\mp \,    & (16) \cr
 &{\bx \nabla} {\cal H}_\pm = \,  {\cal H}_\pm \, {\bx \nabla}
 \, ,
 \qquad \qquad {\cal H}_\pm {\bf r} \, = \, {\bf r} \,
 {\cal H}_\pm \,  \cr
 &( \textstyle{1 \over r} \, {\cal H}_{ e / o } \, r )^{\dag } \,
= \, - \, {1 \over r } \, {\cal H}_{ o / e } \, r \,
 \quad \Rightarrow \qquad  ( {1 \over r } \, {\cal H}_{\pm} \,
 r )^{\dag } \, = \, - \, {1 \over r } \, {\cal H}_{\mp}
 \, r \, .  & (17) \cr
    }  $$
(15) and (17) imply that
$  ( i \, {1 \over r } \, {\cal H}_+ \, r ) \, $ is a
unitary operator.

 Returning to the $ \tilde{p}^0 $ operator we first verify that
 $ {\cal H}_+  \{ \cos ( p^0 \, r )  \,
  e^{ i \, {\bf p} \cdot {\bf r} }  \} \, = \, - \,
  \{ \sin ( p^0 \, r )  \,
  e^{ i \, {\bf p} \cdot {\bf r} }  \} \, ,$ we will
  use ${\cal H}_o \sin (\lambda x) = \cos (\lambda x) \, , \;
  {\cal H}_e \cos (\lambda x) = \, - \, \sin (\lambda x) \, $
  for $ \lambda > 0 \, ,$ also noting that
  $ ( p^0 r + \,  {\bf p} \cdot {\bf r} ) \, > \, 0 \, .$
 $$ \eqalignno{
   {\cal H}_+  \{ \cos ( p^0 r )  \,
  e^{ i \,{\bf p} \cdot {\bf r} }  \} \,
 &\equiv {1 \over 2} \, \left[ \, ( {\cal H}_e + {\cal H}_o )  \,
  + \,  ( {\cal H}_e - {\cal H}_o )  \, {\cal P} \, \right] \,
  \{ \cos ( p^0 r )  \,
  e^{ i \, {\bf p} \cdot {\bf r} }  \} \, \cr
 &= {1 \over 4} \, \left[ \, {\cal H}_e ( 1 + {\cal P} ) \,
  + {\cal H}_o  ( 1 - {\cal P} ) \,  \right] \,
  \{ e^{i ( p^0 r \, + \, {\bf p} \cdot {\bf r} ) }
   \, + \,
  e^{- i ( p^0 r \, - \, {\bf p} \cdot {\bf r} ) }
    \} \, \cr
 &= {1 \over 4} \, \big[ \, {\cal H}_e  \,
  \{ e^{i ( p^0 r \, + \, {\bf p} \cdot {\bf r} ) } \,
  + \, e^{- i ( p^0 r \, - \, {\bf p} \cdot {\bf r} ) }
    \, + \,
  e^{i ( p^0 r \, - \,{\bf p} \cdot {\bf r} ) } \,
  + \, e^{- i ( p^0 r \, + \, {\bf p} \cdot {\bf r} ) }
    \} \, \cr
 &\qquad \qquad + \,  {\cal H}_o  \{
  e^{i ( p^0 r \, + \, {\bf p} \cdot {\bf r} ) } \,
  + \, e^{- i ( p^0 r \, - \,{\bf p} \cdot {\bf r} ) }
    \, - \,
  e^{i ( p^0 r \, - \, {\bf p} \cdot {\bf r})  } \,
  - \, e^{- i ( p^0 r \, + \, {\bf p} \cdot {\bf r} )} \} \, \cr
 &= {1 \over 2} \, \Big[ \, {\cal H}_e
  \{ \cos ( p^0 r + {\bf p} \cdot {\bf r} ) \,
  + \cos ( p^0 r - {\bf p} \cdot {\bf r} ) \} \,
   + \, i \, {\cal H}_o  \{ \sin (p^0 r
 + {\bf p} \cdot {\bf r} ) \, - \, \sin ( p^0 r
  -  {\bf p} \cdot {\bf r} ) \} \, \Big] \, \cr
 &= {1 \over 2} \, \Big[ \, - \, \{
  \sin ( p^0 r \, + \, {\bf p} \cdot {\bf r} ) \,
  + \, \sin ( p^0 r \, - \, {\bf p} \cdot {\bf r} ) \} \,
   + \, i \, \{ \cos ( p^0 r \,
 + \, {\bf p} \cdot {\bf r} ) \, - \cos ( p^0 r \,
  - \, {\bf p} \cdot {\bf r} ) \} \, \Big] \, \cr
 &= {i \over 2} \, \Big[ \,
  e^{i ( p^0 r \, + \, {\bf p} \cdot {\bf r} )} \,
  - e^{ - i (  p^0 r \, - \,{\bf p} \cdot {\bf r} )} \,
   \Big] \, \cr
 &= \, - \, \sin ( p^0 r )  \,
  e^{ i \, {\bf p} \cdot {\bf r} ) }  \, & (18) \cr
    }   $$
    and conversely
    $$ {\cal H}_-  \{ \sin ( p^0 r )  \,
  e^{ i \, {\bf p} \cdot {\bf r}  }  \} \, =  \,
  \{ \cos ( p^0 r )  \,
  e^{ i \,{\bf p} \cdot {\bf r}  }  \} \, .
     \eqno (19) $$
  We substitute this last result into (12) obtaining
$$ \eqalignno{
  (  \textstyle{1 \over 2} \, m^2 c^2 r \,
  + \, a^0 )  \,  \{\sin ( p^0 r )  \,
  e^{ i \, {\bf p} \cdot {\bf r} }  \}
 &= \, - \, ( \partial_r r ) \, p^0 {\cal H}_-
  \{ \sin ( p^0 r )  \,
  e^{ i \, {\bf p} \cdot {\bf r} }  \} \,  \cr
 (  \textstyle{1 \over 2} \, m^2 c^2 r \,
  + \, a^0 )  \, r \; \psi_{\bf v}
 &= \, - \, \partial_r r \, {\cal H}_- r \;
  p^0 \psi_{\bf v} \, = \, - \, i \, \Sigma \,
 {\cal H}_- r \;  p^0 \psi_{\bf v}  \,  & (20) \cr
 \noalign{\noindent now multiply (20) from the left by
 $ - \,i \, {\textstyle{1 \over r}} \, {\cal H}_+
 \Sigma^{- 1} $ to obtain }
 - \, i \, {\textstyle{1 \over r}} \, {\cal H}_+  \Sigma^{- 1}
  ( \textstyle{1 \over 2} \, m^2 c^2 r \,
  + \, a^0 )  \, r \; \psi_{\bf v}
 &= \, p^0 \psi_{\bf v}  \, \cr
  \noalign{\noindent  so that }
  \tilde{p}^0 = \, - \, i \, {\textstyle{1 \over r}} \,
    {\cal H}_+  \Sigma^{- 1}
 (  \textstyle{1 \over 2} \, m^2 c^2 r \, + \, a^0 ) \, r \,
 &= \, - \, {\textstyle{i \over 2}} \, {\textstyle{1 \over r}} \,
  {\cal H}_+  \Sigma^{- 1} r \, (  m^2 c^2 - \, \nabla^2 ) \, r \,
    & (21) \cr
    }  $$
    and we have now found the complete set of velocity operators
    $ ( \tilde{p}^0 , \tilde{\bf p} ) \, .$
 Note that the operator on the LHS of (20), which is
 $ \, {1 \over 2} \, r  \, ( m^2 c^2 - \nabla^2 ) \,
 r \, ,$
 is by inspection a positive operator; and we will later show
 that $ (- \, \partial_r r \, {\cal H}_- r ) \, $ on the RHS
 of (20) is also positive.
  The operators (11) and (21) are the energy-momentum
   4-vector operator $ \tilde{p}^\lambda :$
$$ \eqalignno{
\tilde{p}^\lambda &\equiv \,( \tilde{p}^0 , \, \tilde{\bf p} ) \,
   \equiv \, \Big(  - \, i \, {\textstyle{1 \over r}} \,
    {\cal H}_+  \Sigma^{- 1}
 (  \textstyle{1 \over 2} \, m^2 c^2 r \,
  + \, a^0 ) \, r \,  , \,
    \, \, \textstyle{1 \over r} \, \Sigma^{-1} \,
   ( - \, \textstyle{1 \over 2} \, m^2 c^2 {\bf r} \,
  + \, {\bf a} ) \, r \, \Big) \, & (22) \cr
  {p}^\lambda &\equiv \, (\sqrt{m^2 c^2 + \, p^2 } , \, {\bf p} ) \,
   \cr
   }  $$
  As we recall from (2), the eigenfunctions of
 $ \tilde{p}^\lambda $ contain the same plane wave eigenfunction
  $ e^{ i {\bf p} \cdot {\bf r} } $ of the usual momentum
  operator $ - i {\bx \nabla} \, ,$ except that our
  $ \psi_{\bf v} $ also has the envelope factor
  $ j_0 ( p^0 \bar{r} ) \, .$  We next find a set of boost
 operators $ {\bf K} \, $ and rotation operators $ {\bf J} \, $
  which together with $ \tilde{p}^\lambda $ satisfy
  the Lorentz group commutation relations.

 \beginsection 3 The Lorentz operators

 We establish the Lorentz group operators which must obey the
 commutation relations
  $$\eqalignno{
&[J^a , \, \tilde{p}^b ]
 = \, i \, \epsilon^{abc} \tilde{p}^c \, , \qquad
 [K^a , \, \tilde{p}^0 ]
 = \, i \, \tilde{p}^a \, , \qquad \quad [K^a , \, \tilde{p}^b ]
 = \, i \, \delta^{ab} \tilde{p}^0 \, , & (23) \cr
&[J^a , \, J^b ] = \, i \, \epsilon^{abc} J^c \, , \qquad
 [J^a , \, K^b ]
 = \, i \, \epsilon^{abc} K^c \, , \quad
 [K^a , \, K^b ] = \, - \, i \, \epsilon^{abc} J^c \, & (24) \cr
    }  $$
  with the antisymmetric tensor $ \epsilon^{123} = 1 \, .$
  The boost, rotation operators $ {\bf K} , \, {\bf J} \, $
  such that
  $ \tilde{p}^\lambda $ is a 4-vector are
  $$  {\bf K} = \, - \, {\bx \nabla} \, {\cal H}_- \, r \, , \,
 \qquad {\bf J} = \,  - i \, ({\bf r} \times {\bx \nabla}) \, ,
  \eqno (26) $$
  the rotation operator is as usual, but the boost operator
  necessarily contains the $ {\cal H}_- $ integral operator.
  We first verify that the components of $  {\bf K} \, $
  satisfy the last formula of (24), using the identities (16)
  $$ \eqalignno{
  [ K_1 , \, K_2 ] &=  [ - \, \nabla_1 \,
  {\cal H}_- \, r   , \, - \, \nabla_2 \,
  {\cal H}_- \, r   ] \, \equiv  \, \nabla_1 \,
  {\cal H}_- \, r \, \nabla_2 \,
  {\cal H}_- \, r \, - \, \nabla_2 \, {\cal H}_- \, r \,
  \nabla_1 \, {\cal H}_- \, r \,     \cr
 &=  \nabla_1 \,  {\cal H}_- \, r \, {\cal H}_- \, \nabla_2 \,
  r \, - \, \nabla_2 \, {\cal H}_- \, r \, {\cal H}_- \,
  \nabla_1 \, r \,  =  \nabla_1 \,  {\cal H}_- \,
  {\cal H}_+ \, r \, \nabla_2 \,
  r \, - \, \nabla_2 \, {\cal H}_- \, {\cal H}_+ \, r \,
  \nabla_1 \, r \,     \cr
 &=  \, - \, \nabla_1 \, r \, \nabla_2 \, r \, + \, \nabla_2 \, r \,
  \nabla_1 \, r \, = \, - \, ({\bf r} \times {\bx \nabla})_3
  = \, - \, i \, J_3 \,  \cr
     }  $$
 as stated.
 
 To verify (23) we first separate out the terms
  containing
 $ m^2 $ in the $ ( \tilde{p}^0 , {\bf \tilde p} ) \, $
 operator of (22) as follows:
    $$ \eqalignno{
   ( \tilde{p}^0 , {\bf \tilde p} ) \,
 &= \, {\textstyle{1 \over r}} \Sigma^{- 1} r \, m^2 c^2
  \Big(  i \, {\cal H}_- r , \, {\bf r} \Big) \,
  +  \, {\textstyle{1 \over r}} \Sigma^{- 1} r \,
   \Big( - {\textstyle{i \over r}} {\cal H}_+ a^0 \, r \,  , \,
 \textstyle{1 \over r} \, {\bf a} \, r \, \Big) \, .& (27) \cr
   }  $$
 The Lorentz operators (26)
 commute with $ {1 \over r} \Sigma r \, ,$ so we now
 have to verify that
 both
 $$ \big(  i \, {\cal H}_- r , \, {\bf r} \big) \,
 \qquad \hbox{and} \qquad
 \big( - {\textstyle{i \over r}} {\cal H}_- a^0 \, r \,  , \,
   \textstyle{1 \over r} \, {\bf a} \, r \, \big) \, $$
   are 4-vectors. In the first case we have
    $$ \eqalign{
  [ {\bf K} , \, i \, {\cal H}_- r  ]
  &\equiv  [ - \, {\bx \nabla} \, {\cal H}_- \, r \,   , \,
   i \, {\cal H}_- r   ] \,
   \equiv   - \,  i \, {\bx \nabla} \, {\cal H}_- \, r \,
   {\cal H}_- r  \, + \, i \,
   {\cal H}_- r  {\bx \nabla} \, {\cal H}_- \, r \,    \cr
 &= \, - \,  i \, {\bx \nabla} \,  r \, {\cal H}_+ \,
   {\cal H}_- r  \, + \, i \,
   r  {\bx \nabla} \, {\cal H}_+ {\cal H}_- \, r \,
  =  \,  i \, {\bx \nabla} \,  r \, r  \, - \, i \,
   r  {\bx \nabla} \, r \, = \, i \, {\bf r}  \cr
  \hbox{ and  }  \qquad \qquad
   [ K^a , \,  r^b  ]
   &\equiv  - \, \nabla^a \, {\cal H}_- \, r \, r^b \,
   + r^b \,  \nabla^a \, {\cal H}_- \, r \,
   = - \, \nabla^a \, r^b \, {\cal H}_- \, r \,
   + r^b \,  \nabla^a \, {\cal H}_- \, r \,
   =  \, i \, \delta^{ab} ( i \, {\cal H}_- \, r )
    \, \cr
     }  \eqno (28) $$
     as stated. This result indicates that $ {\bf r} $
     is the space part of a 4-vector: the space-time
     implications of this we will discuss elsewhere. Also
   $$ \eqalign{
  [ {\bf K} , \, - {\textstyle{i \over r}} {\cal H}_+ a^0 \,
  r \,  ]
  &\equiv  [ - \, {\bx \nabla} \, {\cal H}_- \, r \,   , \,
 -  {\textstyle{i \over r}} {\cal H}_+ a^0 \,  r \, ] \,
   \equiv  \,  i \, {\bx \nabla} \, {\cal H}_- \,
    {\cal H}_+ a^0 \,
  r \, \, - \, i \, {\textstyle{i \over r}} {\cal H}_+ a^0 \,
  r \, {\bx \nabla} \,  {\cal H}_- r    \cr
 &= \, - \,  i \, {\bx \nabla} \,a^0 \,
  r \, \, - \, i \, {\textstyle{i \over r}} {\cal H}_+
  {\cal H}_-  a^0 \, r \, {\bx \nabla} \, r  \,
 = \, - \,  i \, {\bx \nabla} \,a^0 \,
  r \, + \, i \, {\textstyle{i \over r}} \,
  a^0 \, r \, {\bx \nabla} \, r    \cr
&= \, {\textstyle{1 \over r}}  [ - \,  i \, r {\bx \nabla} \, ,
 \, a^0 \, ] \, r
  =  i \, {\textstyle{i \over r}} \, {\bf a} \, r \, , \cr
     }  \eqno (29) $$
     the commutator on the last line is a known result (see
     for example [8]).
     To derive (28,29) we have used (16):
 $$ \eqalign{
 &\partial_r {\cal H}_\pm = \,  {\cal H}_\mp \, \partial_r \, ,
 \qquad {\cal H}_\pm r \, = \, r \, {\cal H}_\mp \,
 \qquad {\cal H}_\pm {\bf \hat r} \,
 = \, {\bf \hat r} \, {\cal H}_\mp \, ,
  \qquad {\bx \nabla} {\cal H}_\pm
  = \,  {\cal H}_\pm \, {\bx \nabla} \, ,
 \qquad  {\cal H}_\pm {\bf r} \, = \, {\bf r} \,
 {\cal H}_\pm \, . \cr
    }  $$

  We next investigate the effect of the boost operator
  $ {\bf K} \, $ of (26) on $ \psi_{\bf v} \, .$
  Under an infinitessimal
 boost $ \epsilon \, $ in the $ z \, $ direction the
 wavefunction at time $ t = 0 \,  $ is transformed as
 $ \big[ \psi_{\bf v} \big]_{t=0}  \, \rightarrow  \,
   ( 1 \, + \, i \, \epsilon \, K_3 ) \,
  \big[ \psi_{\bf v} \big]_{t=0}  \, .$ We use (19) to
  obtain
  $$ \eqalignno{
  ( 1 \, + \, i \, \epsilon \, K_3 ) \,
  \big[ \psi_{\bf v} \big]_{t=0}  \,
 &= \, { \sin ( p^0 r )  \over r  } \,
  \exp [  i \,  {\bf p} \cdot {\bf r}] \,
   - \, i \, \epsilon \; \nabla_3 \,
  {\cal H}_- \, r  \Big[ { \sin ( p^0 r )  \over r  } \,
  \exp [  i \, {\bf p} \cdot {\bf r}] \, \Big]   \cr
 &= \, { \sin ( p^0 r )  \over r  } \,
  \exp [  i \, {\bf p} \cdot {\bf r} ] \,
   - \, i \, \epsilon \; \nabla_3 \,
 \Big[  \cos ( p^0 r )  \,
  \exp [  i \,{\bf p} \cdot {\bf r} ] \, \Big]   \cr
 &= \, \Big[ (  1 \, + \,  i \, \epsilon \, p^0 z  ) \,
    { \sin ( p^0 r ) \over r  } \, + \, \epsilon \,
    p_3 \, \cos ( p^0 r ) \Big] \,
   \exp [  i \, m  {\bf v} \cdot {\bf r} ] \, . & (30) \cr
       }  $$
 We can show that this is equivalent to substituting
 $$ p_3 \rightarrow p_3 + \epsilon \, p^0 , \quad
 p_1 \rightarrow p_1 \, , \quad  p_2 \rightarrow p_2 \, ,\quad
 p^0 \rightarrow p^0 + \epsilon p_3           \eqno (31) $$
 into $ \psi_{\bf v} \, ,$ as then
 $$ \eqalignno{
   \big[ \psi_{\bf v} \big]_{t=0}
 &\rightarrow \, { \sin ( (p^0 + \epsilon p_3  ) \, r )
  \over r  } \,
  \exp [  i \, ({\bf p} \cdot {\bf r} \, + \, \epsilon \,
  p^0  z ) ] \,  \cr
 &\approx \, \Big[  { \sin ( p^0  r ) \over r  } \,
  + \, \epsilon \, p_3 \, r \,
  { \cos ( p^0 r ) \over r  } \Big] \,
  \exp [  i \, \epsilon \, p^0  z  ] \,
   \exp [  i \,  {\bf p} \cdot {\bf r} ] \, \cr
 &\approx \, \Big[ (  1 \, + \,  i \, \epsilon \, p^0 z  ) \,
    { \sin ( p^0 r ) \over r  } \, + \, \epsilon \,
    p_3 \, \cos ( p^0 r ) \Big] \,
   \exp [  i \,  {\bf p} \cdot {\bf r} ] \, & (32) \cr
       }  $$
  which is (30). This confirms that $ {\bf K} \, $
  generates the transformation (31).


\beginsection 4 The positive definite scalar product space
  ${\big\langle \phi \, \big| \, \psi \big\rangle}_S  $

 We establish in the Appendix that
 $$ ( - \partial_r r \, {\cal H}_- \, r \, ) \, = \,
 ( - i \, \Sigma \, {\cal H}_- \, r \, ) \,
 \qquad \qquad \hbox{is a positive operator. } \eqno (33) $$
 Recalling (17):
 $$  ( {1 \over r } \, {\cal H}_{\pm} \,  r )^{\dag } \,
 = \, - \, {1 \over r } \, {\cal H}_{\mp} \, r \, , \,  $$
 and also
 $$  \Sigma^{\dag } \, =  \, {1 \over r } \, \Sigma
 \, r \, ,$$
 then
 $$ ( - i \, \Sigma \, {\cal H}_- \, r \, )^{\dag } \,
  \equiv \, \big( - i \, \Sigma \, r \,
  ( \textstyle{1 \over r} \, {\cal H}_- \, r ) \, \big)^{\dag }
  \, = \, \big( - i \, ( \textstyle{1 \over r} \, {\cal H}_+ \, r )
  \Sigma \, r  \, \big) \,
 = \, \big( - i \, \Sigma \, ( \textstyle{1 \over r} \,
  {\cal H}_+ \, r ) \, r  \, \big) \,
 = \, \big( - i \, \Sigma \, {\cal H}_-  r  \, \big) \, , $$
     so that  $  ( - i \, \Sigma \, {\cal H}_- \, r \, ) \, $
  is self-adjoint as well as positive. So we can construct
  a positive definite scalar product space, a Hilbert space:
 $$ {\big\langle \phi \, \big| \, \psi \big\rangle}_S  \,
  \equiv \, {\big\langle \phi \, \big| \,
  - i \, \Sigma \, {\cal H}_- \, r \, \psi \big\rangle} \, .
  \eqno (34) $$
 We will verify that the $ {\bf K} , \, p^\lambda
 $ operators of the last section are self-adjoint with respect
 to  $ {\big\langle \phi \, \big| \, \psi \big\rangle}_S  \, .$
 (The rotation operators commute with the  $  ( - i \, \Sigma \,
 {\cal H}_- \, r \, ) \, $ operator).
 We show the working involved to check the adjoint of
 $ {\bf K} \equiv \, - \, {\bx \nabla} \,
 {\cal H}_- \, r \, :$
 $$ \eqalignno{
  {\big\langle \phi \, \big| \, {\bf K} \, \psi \big\rangle}_S  \,
 &\equiv \, {\big\langle \phi \, \big| \,
  - i \, \Sigma \, {\cal H}_- \, r \, {\bf K} \,
  \psi \big\rangle} \, \equiv \, {\big\langle \phi \, \big| \,
   i \, \Sigma \, {\cal H}_- \, r \, {\bx \nabla} \,
  {\cal H}_- \, r \,  \psi \big\rangle} \, \cr
 &= \, {\big\langle \phi \, \big| \,
  i \, {\cal H}_- \, \Sigma \, r \, {\bx \nabla} \,
  {\cal H}_- \,
  r \,  \psi \big\rangle} \,
  = \, \big\langle - \textstyle{1 \over r^2} {\cal H}_+ r^2 \,
  \phi \, \big| \,
   i \, \Sigma \, r \, {\bx \nabla} \, {\cal H}_- \, r \,
  \psi \big\rangle \, \cr
 &= \, \big\langle - \textstyle{1 \over r^2}
  {\cal H}_+ r^2 \, \phi \, \big| \,
   i \, r \, {\bx \nabla} \, \Sigma \, {\cal H}_- \, r \,
  \psi \big\rangle \,
  = \, \big\langle {\bx \nabla} \, \textstyle{1 \over r}
  {\cal H}_+ r^2 \, \phi \,
  \big| \, i \, \Sigma \, {\cal H}_- \, r \,
  \psi \big\rangle \, \cr
 &= \, \big\langle  {\bx \nabla} \, {\cal H}_- \, r \, \phi \,
  \big| \, i \, \Sigma \, {\cal H}_- \, r \,
  \psi \big\rangle \,  = \, {\big\langle {\bf K} \, \phi \,
  \big| \, \psi \big\rangle}_S \, . \cr
   }  $$
 In going to the last line of working we have used
 $ {\cal H}_\pm \, r \, = \, r \, {\cal H}_\mp \, $ from (16).
 In a similar fashion the self-adjointness
 of the $ p^\lambda $ can be verified with the adjoint property
$$ ( a^\lambda r )^{\dag} = \, ( a^\lambda r ) \, .
 \eqno (35) $$

  \beginsection 5 The inner product spaces
  $ {\big\langle \phi \, \big| \, \psi \big\rangle}_S \, $ and
  $ {\big\langle \phi \, \big| \, \psi \big\rangle} \, .$

 We have established in section 2 that the $ \psi_{\bf v} $
 are eigenvectors of $ \tilde{p}^\lambda $ which is symmetric
  with  respect to the non-negative scalar product space
 $ {\big\langle \phi \,
  \big| \, \psi \big\rangle}_S \, .$ So
 two wavepackets $ \psi_{\bf v} , \, \psi_{\bf v'} $
 momentarily coinciding are necessarily orthogonal:
  $$ {\big\langle \psi_{\bf v} \,
  \big| \, \psi_{\bf v'} \big\rangle}_S = \, 0
  \qquad \quad \hbox{for } {\bf v} \, \neq {\bf v'}\, $$
  In the Appendix we check the
   above result and calculate $ {\big\langle \psi_{-{\bf v}} \,
  \big| \, \psi_{\bf v} \big\rangle}_S $ and
  $ {\big\langle \psi_{{\bf v}} \,
  \big| \, \psi_{\bf v} \big\rangle}_S \, .$
  We find that
     $$ \eqalignno{
  {\big\langle \psi_{\bf v}  \, \big| \,
  \psi_{\bf v'}  \big\rangle}_S
 &= \, ( 2 \, \pi )^3 { p^0 \over 2} \,
  \delta \big( {\bf p} - {\bf p'} )  \,
  \, = \, 4 \, \pi^3  p^0 \delta \big( {\bf p}
  - {\bf p'} )   \, .  & (33)   \cr
    }   $$

  The physical meaning of the scalar product space
  $ {\big\langle \phi \, \big| \, \psi \big\rangle}_S \, $ is
  obscure. More interesting is the usual
  scalar product of two wavefunctions
  $ {\big\langle \phi \, \big| \, \psi \big\rangle} \, ,$
  the $( \phi^* \psi ) \, $ within the integral being a
  conserved quantity for any $ \phi , \,  \psi  \, $
  satisfying (1). Consider two wavepackets $ \psi_v , \,
  \psi_{v'} $ with velocities $ v , \, v' $ in the $ z \, $
 direction which momentarily coincide at time  $ t = 0 \, ,$
 so that $ \psi_v = \sin ( p^0 r )  \,
  e^{  i \, p \, z  } / r \, $ and $ \psi_{v'}
  = \sin ( p'^0 \, r )  \, e^{  i \, p' \, z  } / r \, ,$ with
  $ - \infty < ( p , p' ) < \infty \, .$
 Then
  $$ \eqalignno{
  {\big\langle \psi_{v'}   \, \big| \, \psi_v  \big\rangle}
 &= \, \int \big( \sin ( p'^0 \, r )  \,
  e^{-  i \, p'\, z } / \, r  \big) \, \big( \sin ( p^0 \, r ) \,
  e^{ i \, p \, z } / \, r \big) \, d^3 {\bf r} \,   \cr
 &= \,  2  \pi \, \int  \sin ( p'^0  r ) \, \sin ( p^0 r ) \;
  e^{  i \, ( p - p' ) r \, \cos \theta  } \;
  d(- \cos \theta ) \, dr  \,   \cr
 &= \,  \pi \,  \int  \Big[ \cos (( p^0 - p'^0 ) r ) \, - \,
  \cos ( ( p^0 + p'^0 ) r ) \, \Big] \,
  \Big[ 2 \, { \sin ( ( p - p' ) r ) \over
  ( p - p' ) \, r  } \Big] \; dr \, \cr
 &= \, {  \pi \over ( p - p' ) } \int  \Big[
 \sin ( [ p - p' + (p^0 - p'^0 ) ] r ) \,
 + \, \sin ( [ p - p' - (p^0 - p'^0 ) ] r ) \, \cr
 & \qquad \quad  \qquad \quad \qquad
 \, - \, \sin ( [ p^0 + p'^0 + p - p' ] r ) \,
  + \, \sin ( [ p^0 + p'^0 - p + p' ] r ) \,    \Big] \,
   \, {1 \over r} \; dr \, \cr
 &= \,  { \pi \over ( p - p' ) } \, {\pi \over 2} \,
   \Big[
 \hbox{sgn} ( p - p' ) \, + \, \hbox{sgn} ( p - p' ) \,
 - \, 1 \, + \, 1 \Big] \, \cr
 &= \, { \pi^2 \over | (p - p' ) | } \, \;  .
   &  (34) \cr
      }   $$
  So that the scalar product of two wavepackets which
  momentarily coincide have an inner product
  $ {\big\langle \psi_{v'} \, \big| \, \psi_v \big\rangle} \, $
  which is a finite number (when $ v \neq
  v' \, ),$ and furthermore this inner product is constant
  for all time, however far apart the wavepackets are
  separated (either before or after the wavepackets have
  coincided). This linkage between two coinciding wavepackets
  or particles is a well known quantum feature.
For two wavepackets coinciding with momenta $ {\bf p} , \,
  {\bf p'} \, $ not necessarily collinear, the above result (34)
  is
  $$ \eqalignno{
  {\big\langle \psi_{\bf v'}   \, \big| \, \psi_{\bf v}
  \big\rangle}
 &= \, {  \pi^2 \over | {\bf p} - {\bf p'}  | } \, \;  .
 &  (35) \cr
      }   $$

  \beginsection 6 Towards a fully relativistic theory

  The evolution equation for the $ \psi_{\bf v} $ is the
  non-relativistic
  Schrodinger equation, and the momentum $ {\bf p} = \,
  m {\bf v} \, $ which implies an unbounded velocity.
  On the other
  hand the energy $ c p^0 = \, c \sqrt{ m^2 c^2 + p^2 } \, $
  as in the relativistic case. It is straightforward to obtain a
  $ \Psi_{\bf v} $ so that the 4-momentum has the
  relativistic value
  $$ ( \tilde{p}^0 , \, {\bf \tilde p} ) \, \Psi_{\bf v} = \,
  ( m \gamma c , \, m \gamma {\bf v} ) \, \Psi_{\bf v} \, ,
  \qquad \qquad \gamma \equiv \,
   ( 1 - v^2 / c^2 )^{- 1 / 2 }   \, ,$$
  but more difficult to find the evolution equation for this
   $ \Psi_{\bf v} .$

   Consider
  $$ \eqalignno{
 \Psi_{{\bf v}}
 &= \, { \sin ( m \gamma \, c | {\bf r} \, - \, {\bf v} t | )
  \over  | {\bf r} \, - \, {\bf v} t |  } \,
  \exp [  i \, m  \gamma {\bf v} \cdot ({\bf r} \,
  - \, {\bf v} t ) ] \; e^{ - \, i \, m c^2 t / 2 \gamma }
  & (36) \cr   } $$
     then from the results of section 2 by simple substitution
   $$   ( \tilde{p}^0 , \, {\bf \tilde p} ) \,
   \big[ \Psi_{\bf v} \big]_{t=0}
   =  ( m \gamma c  , \, m \gamma {\bf v} \, )  \,
   \big[ \Psi_{\bf v} \big]_{t=0}  \eqno (37) $$
  and now the $ \Psi_{{\bf v}} \, $ (while still having velocity
  $ {\bf v} \, )$ has
   the correct relativistic 4-momentum when acted on
     by the $ \tilde{p}^\lambda $ operator. We look for the
 evolution equation for $ \Psi_{{\bf v}} \, .$ Proceeding as in
 section 1, we note the identities
  $$ \eqalignno{
  &( \partial_t + {\bf v} \cdot {\bx \nabla} \,
   + { i \, m c^2 \over 2 \gamma } ) \,
\Psi_{{\bf v}} \, = \, 0 & (38) \cr
&( {\bx \nabla } \, - \, i \, m \, \gamma \, {\bf v} ) \,
 \Psi_{\bf v} \,
 = \, e^{  i \, m  {\bf v} \cdot ( {\bf r} \,
 - \, {\bf v} t ) } \, e^{ - \, i \, m c^2 t / 2 \gamma  } \; \,
  {\bx \nabla } \, \Big[{ \sin ( m \gamma | {\bf r} \,
  - \, {\bf v} t | )
  \over  | {\bf r} \, - \, {\bf v} t |  } \Big] \,  \cr
 &( {\bx \nabla } \, - \, i \, m \, \gamma \, {\bf v} )^2 \,
   \Psi_{\bf v} \, = \, e^{  i \, m  {\bf v} \cdot ( {\bf r} \,
  - \, {\bf v} t ) } \, e^{ - \, i \, m c^2 t / 2 \gamma } \; \,
  \nabla^2 \, \Big[{ \sin ( m \gamma | {\bf r} \,
  - \, {\bf v} t | )
\over  | {\bf r} \, - \, {\bf v} t |  } \Big]  \,
 = \, - \, m^2 \gamma^2 c^2 \, \Psi_{\bf v} \, . & (39)  \cr
   }  $$
Expanding out (39)
 and substituting in (38) we obtain
   $$ \eqalignno{
  &\Big( \nabla^2 \, - \, 2 \, i \, m \, \gamma \,
   {\bf v} \cdot {\bx \nabla} \,
   - \, m^2 \gamma^2 v^2 \Big)  \, \Psi_{\bf v}  \,
   = \, - \, m^2 \gamma^2 c^2 \, \Psi_{\bf v} \cr
  &\Big( \nabla^2 \, + \, 2 \, i \, m \, \gamma \,
   ( \partial_t  + { i \, m c^2 \over 2 \gamma } ) \, \Big) \,
    \Psi_{\bf v} \,  = \, - \, m^2 c^2 \, \Psi_{\bf v} \cr
  &\Big( \nabla^2 \, + \, 2 \, i \, m \, \gamma
   \, \partial_t \Big) \,  \Psi_{\bf v} \,  = \,
   \Big( \nabla^2 \, + \, 2 \, i \, {p^0 \over c}
   \, \partial_t \Big) \,  \Psi_{\bf v} \,  = \, 0 & (40) \cr
   }  $$
   which apart the gamma factor in the partial time
   derivative is just the Schrodinger equation (1). But
   due to the $ \gamma \, $ factor which depends on the
  wavepacket velocity, equation (40) is not a true evolution
  equation applicable to any $ \Psi_{\bf v} .$
   We can try inserting the
   $ \tilde{p}^0 $ operator instead of $ p^0 $ in (40), but
   the $ \tilde{p}^0 $ operator only has the eigenvalue
   property when the origin is at the wavepacket centre, which
   is not desirable for a evolution equation
   applicable wherever the origin.

 \beginsection 7 Outlook

  The attraction of the wavepacket solutions
  $ \psi_{\bf v} \, $
  is their localised nature which accords with our
  everyday experience. We have made progress in showing
that the $ \psi_{\bf v} \, $ can be used as alternative basis
states to the plane waves, albeit with some complications: for
example the momentum operators (22) only have the eigenfunction
 property when the origin is at the wavepacket centre.

The usual momentum operator $ {\bmit \pi} \equiv
- i \, {\bx \nabla } \, $ does not commute with the momentum
operators $ \tilde{\bf p} $: the wavepackets are not invariant
under translation, unlike a plane wave - which merely undergoes
a phase change on translation. (Of course the
$ \psi_{\bf v} \, $ are invariant under a combined space and
time translation.)
 The $ \psi_{\bf v} \, $
are not eigenfunctions of the Hamiltonian operator $ -
\, \nabla^2 / 2 m \, :$ it is impossible to have stationary
states of the form $ \psi = e^{- i E t} f({\bf r}) \, $
 for wavepackets in motion, as the wavepacket envelope must
itself be a function of $ t \, $.
An unexpected feature of the wavepacket momentum operator
 is that $ \tilde{\bf p} \, $ is part of a 4-vector with
 $ m^2 c^2 + \, ( \tilde{\bf p} )^2 = \, ( \tilde{p}^0 )^2 \, ,$
 as discussed in section 3. From (30) we see how the boost
 operator generates a change in momentum. However the
velocity has no upper limit,
 the operator $ \tilde{\bf p} / m \, $ is unbounded. In a
 fully relativistic theory we would expect the wavepacket
 velocity to be less than $ c \, ;$ so far, as shown in the
 previous section, we have had limited success in the
 relativistic wavepacket extension. We have shown
 that the position $ {\bf r} \, $ is part of a 4-vector
 with $ r^0 = \, i \, {\cal H}_- r \, ,$ which has
 far-reaching implications on the nature of space-time.
 We will discuss this further elsewhere.

 Operators very similar to
$ \tilde{\bf p} \, $ have been discussed before in the context
of lightcone quantum mechanics, an approach initiated by Dirac
[10] in 1949 and subsequently worked on by a number of authors
[7,8,9]. Their approach was to start
with the Poisson bracket formulation of classical mechanics,
find the lightcone generators, and then `quantise'
these generators to appropriate operators. The second of these
authors [8] found a canonical transformation from the usual
conjugate pair $ ({\bf r} , \, {\bmit \pi} ) \, \rightarrow  \, (
{\bf x} , \, {\bf p} ) \, , $  where
$$ {\bf p}  ( {\bf r} \, , \, {\bmit \pi} )
  =  \, {\bmit \pi} \, - \, { m^2 c^2 +  \pi^2 \over
 2 \, ( {\bf r} \cdot {\bmit \pi} ) } \, {\bf r}  \, ,
 \qquad \qquad {\bf x}  ( {\bf r} \, , \, {\bmit \pi} )
  = \, { 2 \, ( {\bf r} \cdot {\bmit \pi} ) \over m^2 c^2 + \pi^2 }
  \, {\bmit \pi}  \, .$$
 Our $\tilde{\bf p} \, $ operator appears to be the quantised
 version of the generator above with
 ${\bmit \pi} \rightarrow - \, i \, {\bx \nabla } \, ,$
 which is more easily seen
if we write $ \tilde{\bf p} \, $ in the form
  $$ \eqalignno{
 \tilde{\bf p} \,
  &=   - \, i \, \textstyle{1 \over r} \, {\bx \nabla } \,
  r \, - \, \textstyle{1 \over 2} \, \textstyle{1 \over r} \,
  \Sigma^{-1} \, {\bf r} \, ( m^2 c^2 - \nabla^2 ) \, r \, ,   \cr
     } $$
     the $ \Sigma \, $ corresponding to the classical
     $ ( {\bf r} \cdot {\bmit \pi}   ) \, .$
  Consider a positive energy plane wave
  $ e^{ - i c p^0 t + i {\bf p} \cdot {\bf r} } \, .$
  On the past lightcone $ t = - r / c \, $
  this becomes $ e^{ i p^0 r + i {\bf p} \cdot {\bf r} } \, ,$
  which referring to (5) is one half of the
  $ \psi_{\bf v} \, $ wavefunction, the other half containing
  $ e^{ - i p^0 r + i {\bf p} \cdot {\bf r} } \, $ can be
  thought of as a future lightcone wave.
  From this point of view the $ \psi_{\bf v} \, $ is a balance
  of past and future lightcone waves.
  The wavepacket approach hints at a more dynamical picture
  of a particle - existing as a confluence of past
  and future lightcone waves.

  \vskip 0.2 in

  \noindent {\bf Acknowledgement }  I am grateful to Prof G
  Kaiser [11] for pointing out that puting the light
  velocity $ c = 1 \, $ as in an earlier version of this
  paper leads to incorrect dimension.

\beginsection  Appendix

 \noindent {\bf  I. The positive operator $ ( - \partial_r r \,
 {\cal H}_- \, r \, ) \, $  } \hfill\break
 \noindent
That $ ( - \partial_r r \, {\cal H}_- \, r \, ) \, $
 is a positive operator is a key result in
this paper, leading to the Hilbert space of
section 4. We reproduce the derivation from ref [6] below.

Consider the unitary operator $ {\cal U} \, $
$$ {\cal U} \, f ({\bf r}) \, \equiv {1 \over 2 r } \,
\big[ ( {\cal F}_c \, - \, i \, {\cal F}_s \, ) \,
+ \, ( {\cal F}_c \, + \, i \, {\cal F}_s \, ) \, {\cal P} \,
\big] \, r \, \, f ({\bf r})  \eqno (A1) $$
 where $ {\cal P } \, $ is the parity operator and
$ {\cal F}_c , \, {\cal F}_s \, $ are the Fourier cosine,
sine transforms defined by
$$ \eqalign{
 {\cal P} \,  f(r , \theta , \phi ) &=  f(r , \pi - \theta ,
 \phi + \pi ) \cr
{\cal F}_c \, f(r , \theta , \phi ) &= g_c (r , \theta , \phi )
\equiv \, \sqrt{2 \over \pi} \,
\int_0^\infty  f(t , \theta , \phi ) \, \cos (r t ) \, dt \, , \cr
{\cal F}_s \, f(r , \theta , \phi ) &= g_s (r , \theta , \phi )
\equiv \, \sqrt{2 \over \pi} \,
\int_0^\infty  f(t , \theta , \phi ) \, \sin (r t ) \, dt \, . \cr
  }    \eqno (A2) $$
The operator $ {\cal U} \, $ has inverse $ {\cal U}^{- 1} $
where
$$ \eqalign{
 {\cal U}^{- 1} = \, {\cal U}^* \,
 &= {1 \over 2 r } \, \big[ ( {\cal F}_c \,
  + \, i \, {\cal F}_s \, ) \, + \, ( {\cal F}_c \,
  - \, i \, {\cal F}_s \, ) \, {\cal P} \, \big]
  \, r \, .   \cr
     } \eqno (A3)  $$
 It is straightforward to check that
$ {\cal U}^* \, {\cal U} \, = 1 \, ,$ with the aid of
$$ {\cal F}_c \, {\cal F}_c   = {\cal F}_s \, {\cal F}_s
= {\cal P} \, {\cal P} \, = 1 \, . \eqno (A4) $$
We emphasise that
$ {\cal U} , \, {\cal U}^* \, $ are in a sense one dimensional
transform operators:
 the integration is along the axis $ \lambda \hat{\bf r} \, ,$
 with $ - \infty > \lambda > \infty \, $.

 We next calculate the operator
 $$ l^0 \equiv \,
  {\cal U}^* \, r \, {\cal U} \,  \eqno (A5) $$
  which is clearly a non-negative operator.
 To simplify (A5), it is known that the cosine transform of the
 sine transform is the Hilbert transform of odd functions,
 and the sine transform of the
 cosine transform is minus the Hilbert transform of even
 functions, i.e.
$$ \eqalign{
 {\cal F}_s {\cal F}_c \, = \, - \, {\cal H}_e \, , \qquad
 {\cal F}_c {\cal F}_s \, = \, {\cal H}_o \, \cr
  } \eqno (A6)  $$
where $ {\cal H}_e , \, {\cal H}_o \, $ are defined by
$$ \eqalign{
{\cal H}_e \, f(r , \theta , \phi ) &\equiv \, - \, { 2 r \over \pi } \,
\int_0^\infty { f(t , \theta , \phi ) \over r^2 - t^2 } \, dt \, , \qquad
 \quad \,{\cal H}_e \, f({\bf r}) = \, - \, { 2  \over \pi } \,
\int_0^\infty { \, f( \lambda {\bf r} ) \over 1 - \lambda^2 } \,
d\lambda \, , \cr
{\cal H}_o \, f(r , \theta , \phi ) &\equiv  \, - \,  { 2 \over \pi } \,
\int_0^\infty { t \, f(t , \theta , \phi ) \over r^2 - t^2 } \, dt \, ,
\qquad \quad
{\cal H}_o \, f({\bf r}) =  \, - \,  { 2 \over \pi } \,
\int_0^\infty { \lambda \, f( \lambda {\bf r} ) \over 1 - \lambda^2 } \,
d\lambda \, . \cr }
  \eqno (A7)   $$
For brevity we adopt the notation
$$  {\cal F}_\pm \equiv {\cal F}_c \, \pm \, i \, {\cal F}_s \, $$
then from (A4,A6) there follows the further identities
$$ \eqalign{
&{\cal F}_+ \, {\cal F}_+
= \, - \, i \, ( {\cal H}_e - {\cal H}_o )  \, , \qquad \qquad
{\cal F}_- \, {\cal F}_-
= \, i \, ( {\cal H}_e - {\cal H}_o )  \, , \qquad  \cr
&{\cal F}_+ \, {\cal F}_-
= 2 \, - \, i \, ( {\cal H}_e + {\cal H}_o )  \, , \quad \qquad
{\cal F}_- \, {\cal F}_+
= 2 \, + \, i \, ( {\cal H}_e + {\cal H}_o )  \, , \qquad \cr
 } \eqno (A8) $$
Also we note that
$$ \eqalignno{
&{\cal F}_\pm \, r
= \mp \, i \, \partial_r \, {\cal F}_\pm  \, ,  \qquad
\qquad  \qquad
 r \, {\cal F}_\pm \,
= \pm \, i \, {\cal F}_\pm  \, \partial_r \, . \qquad
\qquad\qquad \quad    & (A9) \cr
 }       $$
Returning to $ l^0 \, = \, {\cal U}^* \,  r \, {\cal U} \, $
we have
$$ \eqalignno{
l^0 \, = \, {\cal U}^* \,  r \, {\cal U} \,
&=  {1 \over 4 r } \,
\left( {\cal F}_+ \, + \, {\cal F}_-  {\cal P} \,
 \right) \, r \, \left( {\cal F}_-  \, + \, {\cal F}_+  {\cal P} \,
 \right) \, r \, \cr
&= \, i \, {1 \over 4 r } \, \partial_r \,
\left( - \, {\cal F}_+  \, + \, {\cal F}_- {\cal P} \, \right) \,
\left( {\cal F}_-  \, + \, {\cal F}_+  {\cal P} \,  \right) \,
 r \, \cr
&= \, i \, {1 \over 4 r } \, \partial_r \,
\left[ ( - \, {\cal F}_+ {\cal F}_-  \,
+ \, {\cal F}_- {\cal F}_+  ) \,
+ \, ( \, {\cal F}_- {\cal F}_-
- \, {\cal F}_+ {\cal F}_+   ) \, {\cal P} \, \right] \, r  \cr
&= \, - \, {1 \over 2 r } \, \partial_r \,
\left[ \, ( {\cal H}_e + {\cal H}_o )  \,
+ \,  ( {\cal H}_e - {\cal H}_o )  \, {\cal P} \, \right]  \,
 r  \,
 \equiv  \, - \, {1 \over r } \, \partial_r \, {\cal H}_+  r \,
      & (A10) \cr
   }  $$
 recalling
$ {\cal H}_+ \equiv \, {1 \over 2} \, [ \, ( {\cal H}_e + {\cal H}_o )  \,
 + \, ( {\cal H}_e - {\cal H}_o )  \, {\cal P} \, ] \,  $
  from (13) which is a Hilbert transform along the axis
$  \lambda \, \hat{\bf r} \, ,\; ( - \infty < \lambda < \infty ) \, .$

The operator components $ \partial_r \, , \, {\cal H}_+ \, $
of $ l^0 \, $ do not commute:
If in the working out of (A10) we take $ r \, $ to the right
instead of to the left, we arrive at
$$ \eqalignno{
l^0 \, = \, {\cal U}^* \,  r \, {\cal U} \,
&= \, - \, {1 \over 2 r } \,
\left[ \, ( {\cal H}_e + {\cal H}_o )  \,
- \,   ( {\cal H}_e - {\cal H}_o )  \, {\cal P} \, \right]  \,
\partial_r \, r \, \cr
&\equiv \, - \, {1 \over r } \, {\cal H}_- \partial_r \,  r \, .
   & (A11) \cr  }  $$
As
$$ {\cal H}_- {\cal H}_+ = {\cal H}_+ {\cal H}_- = \, - \, 1 \, ,$$
which can be verified their definitions (13) together
 with
 $ {\cal H}_e {\cal H}_o = {\cal H}_o {\cal H}_e
 = \, - \, 1 \, ,$ then
$$ {l^0}^2
 = \, {1 \over r } \, \partial_r \, {\cal H}_+
 {\cal H}_-  \partial_r \,  r \,
 = \, - {1 \over r } \, \partial_r^2 \,  r \, $$
 as expected. Hence $ l^0 $ is a positive square root of
  $ \, - {1 \over r } \, \partial_r^2 \,  r \, .$

  Finally from (A10), (A11)
   $$ \, r \, l^0 \, r \, = \, ( - \partial_r {\cal H}_+ \,
   r^2  ) \, = \, ( - \partial_r r \,
  {\cal H}_- \, r \, ) \, \eqno (A12) $$
  and as we have shown $ l^0 $ is non-negative, so must be the
  operator $ r l^0 r = \, ( - \partial_r r \, {\cal H}_- \,
  r \, ) \, .$

          \vskip 0.1 in

 \noindent {\bf II. To calculate $ {\big\langle \psi_{-{\bf v}} \,
  \big| \, \psi_{\bf v} \big\rangle}_S \, $  } \hfill\break
 \noindent
  We calculate $ {\big\langle \psi_{\bf v} \,
  \big| \, \psi_{\bf v'} \big\rangle}_S \, ,$ the inner product
  of two wavepackets coinciding momentarily at time
  $ t = 0 \, .$ If $ {\bf v} \neq {\bf v'} \, $ then there
  will be a centre-of-momentum frame where $ {\bf v}
  = \, - \, {\bf v'} \, .$ (The exponentiated boost
  operator $ e^{{\bf c}\cdot{\bf K} } $ can be applied
  to the $ \psi_{\bf v} , \psi_{\bf v'}  $
  to arrive at the
  centre-of-momentum frame.) Before we calculate
  $ {\big\langle \psi_{ - {\bf v} }  \, \big| \,
  \psi_{\bf v}  \big\rangle}_S \, $
  we first simplify the inner product
  $ {\big\langle \phi \, \big| \, \psi \big\rangle}_S \, $ :
$$ \eqalignno{
  {\big\langle \phi \, \big| \, \psi \big\rangle}_S
 &\equiv {\big\langle \phi \, \big| \, ( - i \, \Sigma \,
 {\cal H}_- \, r \, ) \, \psi \big\rangle}
  =  {\big\langle \phi \, \big| \, - \partial_r r \,
 {\cal H}_- \, r  \, \psi \big\rangle} \cr
 &= {\textstyle{1 \over 2}} \,
  {\big\langle \, {\textstyle{1 \over r}}\, \partial_r r^2 \,
 \phi \, \big| \, {\cal H}_- \, r  \, \psi \big\rangle} \,
 + \,  {\textstyle{1 \over 2}} \, {\big\langle \phi \, \big| \, - \partial_r r \,
 {\cal H}_- \, r  \, \psi \big\rangle}  \cr
 &= {\textstyle{1 \over 2}} \,
  {\big\langle \, (\partial_r + \, {\textstyle{1 \over r}}\, )
  \, r \, \phi \, \big| \, {\cal H}_- \, r  \, \psi \big\rangle} \,
 + \,  {\textstyle{1 \over 2}} \, {\big\langle \phi \, \big| \,
  - \, r (\partial_r + \, {\textstyle{1 \over r}}\, ) \,
 {\cal H}_- \, r  \, \psi \big\rangle}  \cr
 &= {\textstyle{1 \over 2}} \,
  {\big\langle \, \partial_r  \, ( r \, \phi ) \, \big| \,
  {\cal H}_- \, r  \, \psi \big\rangle} \,
  + \,  {\textstyle{1 \over 2}} \, {\big\langle r \, \phi \,
  \big| \, - \, \partial_r ( {\cal H}_- \, r  \, \psi ) \,
  \big\rangle}  \cr
 &\equiv \, {\textstyle{1 \over 2}} \,
  \int \Big[ \big( \partial_r  \,
  ( r \, \phi ^* ) \big) \, ( {\cal H}_- \, r  \, \psi ) \, + \,
 \big(   r \, \phi ^*  \big) \, ( - \, \partial_r  \,
  ( {\cal H}_- \, r  \, \psi ) \,\big) \, \Big] \, \,
   d^3 {\bf r} \,  & (A13)  \cr
    }   $$
 Now we substitute
  $ \psi = \psi_{\bf v} , \phi = \psi_{- {\bf v}} \,  $
  into the above, recalling from (19) that
  $ {\cal H}_- r \psi_{\bf v} \, =  \,
  \{ \cos ( p^0 r )  \,
  e^{ i \, {\bf p} \cdot {\bf r} }  \} \, ,$ obtaining
  $$ \eqalignno{
  {\big\langle \psi_{ - {\bf v} }  \, \big| \,
  \psi_{\bf v}  \big\rangle}_S
 &= \, {\textstyle{1 \over 2}} \, \int \Big[
  \partial_r \big( \sin ( p^0 r )  \,
  e^{ i \, {\bf p} \cdot {\bf r} } \big) \,
  ( \cos ( p^0 r )  \,
  e^{ i \, {\bf p} \cdot {\bf r} } ) \, \cr
  &\qquad \qquad \qquad \qquad  - \,
   \big( \sin ( p^0 r )  \,
  e^{ i \, {\bf p} \cdot {\bf r} } \big) \,
  ( \partial_r  \cos ( p^0 r )  \,
  e^{ i \, {\bf p} \cdot {\bf r} } ) \, \Big]
  d^3 {\bf r} \,   \cr
 &= \, { p^0 \over 2} \, \int \Big[
   \big( \cos ( p^0 r )  \, + \, i \,
   ( {{\bf p} \over p^0} \cdot \hat {\bf r} ) \, \sin ( p^0 r )
   \big) \, \cos ( p^0 r )   \, \cr
  &\qquad \qquad \qquad \qquad  + \,
    \sin ( p^0 r )  \,
  \big( \sin ( p^0 r )  \, - \, i \,
   ( {{\bf p} \over p^0} \cdot \hat {\bf r} ) \, \cos ( p^0 r )
   \big) \, \Big] \,
  e^{ 2\, i \, {\bf p} \cdot {\bf r}  } \, d^3 {\bf r} \, \cr
 &= \, { p^0 \over 2} \, \int
  e^{ 2\, i \, {\bf p} \cdot {\bf r} ) }  \,
  d^3 {\bf r} \, = \, ( 2 \, \pi )^3 { p^0 \over 2} \,
  \delta \big( {\bf p} - ( - {\bf p} ) \big) \,
  = \, 0 \, \qquad \hbox{for } \quad {\bf p} \neq \, 0 \, .
     &  (A14) \cr  }   $$
  If instead we substitute $ \psi =  \phi = \psi_{\bf v} \,  $
  into (30) we obtain
  $$ \eqalignno{
  {\big\langle \psi_{\bf v}  \, \big| \,
  \psi_{\bf v}  \big\rangle}_S
 &= \, { p^0 \over 2} \, \int \Big[
   \big( \cos ( p^0 r )  \, - \, i \,
   ( {{\bf p} \over p^0} \cdot \hat {\bf r} ) \, \sin ( p^0 r )
   \big) \, \cos ( p^0 r )   \, \cr
  &\qquad \qquad \qquad \qquad  + \, \sin ( p^0 r )  \,
  \big( \sin ( p^0 r )  \, - \, i \,
   ( {{\bf p} \over p^0} \cdot \hat {\bf r} ) \, \cos ( p^0 r )
   \big) \, \Big] \,  d^3 {\bf r} \,   \cr
 &= \, { p^0 \over 2} \, \int
  \big[ 1 \, - 2 \, i \, ( {{\bf p} \over p^0} \cdot
  \hat {\bf r} ) \, \sin ( p^0 r ) \cos ( p^0 r ) \, \big] \,
    d^3 {\bf r} \,
   = \, ( 2 \, \pi )^3 \, { p^0 \over 2} \,
   \delta \big( {\bf p} - {\bf p} ) \big) \,
 \, , & (A15) \cr
    }   $$
   as integrating over the angular variable
    $ ( \hat{\bf p} \cdot \hat{\bf r} )  = \cos \theta \, $
    annihilates the second term in the integral.
    We conclude that
     $$ \eqalignno{
  {\big\langle \psi_{\bf v}  \, \big| \,
  \psi_{\bf v'}  \big\rangle}_S
 &= \, ( 2 \, \pi )^3 { p^0 \over 2} \,
  \delta \big( {\bf p} - {\bf p'} )  \,
  \, = \, 4 \, \pi^3  p^0 \delta \big( {\bf p}
  - {\bf p'} )   \, .  & (A16)   \cr
    }   $$

                   \vskip 0.1 in

\beginsection References

\item{[1]} Holland P R, 1994 {\it The Quantum Theory of Motion}
(Cambridge University Press, Cambridge)
\item{[2]} Besieres I M et al, 1994 {\it Am. J. Phys. }
{\bf 62} 519
\item{[3]} Barut A O, 1990 {\it Found. Phys. } {\bf 20} 1233
\item{[4]} Berry M V and Balazs N L, 1979 {\it Am. J. Phys. }
{\bf 47} 264
\item{[5]} Mosley S N, 1996 {\it J. Phys. A } {\bf 29} 6671
\item{[6]} Mosley S N, 2003 {\it arXiv}: quant-ph0310159
\item{[7]} Peres A, 1967 {\it J. Math. Phys. } {\bf 98} 785
\item{[8]} Derrick G H, 1987 {\it J. Math. Phys. } {\bf 28} 1327
\item{[9]} Mosley S N and Farina J E G, 1992
{\it J. Phys. A } {\bf 25} 4673
\item{[10]} Dirac P A M, 1949 {\it Rev. Mod. Phys. } {\bf 21} 392
\item{[11} Kaiser G, 2008 (\it private communication)

\end